# Lineshapes in coherent anti-Stokes Raman spectra of gases

**Michele Marrocco,**[1,*] **Emil Nordström,**[2] **Per-Erik Bengtsson,**[2]

[1]*ENEA, via Anguillarese 301, 00167 Rome, Italy*
[2]*Combustion Physics, Lund University, SE-22100, Sweden*
*\*Corresponding author: michele.marrocco@enea.it*

We introduce new lineshapes of coherent anti-Stokes Raman spectra of gases by considering realistic nanosecond laser pulses that deviate from the Gaussian representations of their time and spectral envelopes. The analysis suggests that the contribution to the linewidth caused by the interaction with such laser pulses could be erroneously attributed to the intrinsic Raman width if the customary CARS approach, based primarily on the Voigt lineshape, is assumed.

Coherent anti-Stokes Raman scattering (CARS) is one of the fundamental phenomena of nonlinear optics [1-4]. It involves the use of three laser fields and, among several applications, gas analysis with nanosecond lasers has attracted a lot of interest over the years [3, 5-8]. Within this context, the diagnostic capacity depends critically on the physical understanding of CARS spectra and, for this reason, one has to secure the information contained in the third-order nonlinear susceptibility describing the emerge of an isolated CARS line. This optical response function is very-well known and is usually represented as a fourth-rank tensor of the kind

$$\chi^{(3)}_{ijkl}(\Delta) = \frac{K_{ijkl}}{\Delta - i\Gamma/2} \quad (1)$$

where $K_{ijkl}$ contains the information about the Raman molecule (Raman frequency, Boltzmann distribution of the molecular population, polarizability derivatives and other molecular parameters), $\Delta$ is the Raman detuning and $\Gamma$ is the spontaneous Raman FWHM linewidth whose value is of paramount importance. It is then tempting to conclude that Eq. (1) leads to Lorentzian profiles in CARS spectra where single Raman modes are isolated.

The conclusion is, however, subject to constraints. In particular, disregarding the interference caused by the non-resonant background, the deviation of the CARS lineshape from the true Lorentzian profile of Eq. (1) has been discussed by several authors with reference to the use of multi-mode lasers [9-13]. On the other hand, the advent of the injection-seeding (or pulsed injection locking) technology [14-17] has made possible to realize single-mode Q-switched laser operation resulting in near-transform-limited pulses with spectral bandwidths $\delta\tilde{v}$ that are sufficiently smaller than the typical Raman widths $\Gamma$. However, irrespective of whether the laser is operated single- or multi-mode, the reference pulse assumed in most of the CARS studies has a Gaussian spectral envelope and this results in the traditional Voigt lineshape obtained as real part of the complex error function [9, 13, 18-20]. Beyond its undoubted importance in laser spectroscopy at large, this function is at the core of various fitting codes developed to study numerically the complex fine structures of CARS spectra. The most known is the so-called CARSFT code [21], but others have been published [20, 22].

In this Letter, we demonstrate that realistic aspects of the laser pulses could determine a significant change in the CARS lineshape in comparison to the Lorentz and, most importantly, Voigt profiles. For instance, Gaussian functions to approximate envelopes of the electric fields of transform-limited single-mode lasers contrast against the time asymmetry of such laser systems [15-17, 23-27]. It is then instructive to see what kind of spectral consequences could be expected if elements of reality are incorporated in the traditional CARS theory.

The main proof is based on some given spectral envelopes of the electric fields whose spectral intensities are assumed to have the same FWHM of 0.005 cm$^{-1}$. The condition refers to known linewidth values of single-mode Nd:YAG lasers and is represented in the lower panel of Fig. 1 where the spectral envelopes are plotted according to their transform-limited pulses reported in the upper panel of Fig. 1. The latter are chosen as follows. First of all, we select two symmetric pulses. One is the Gaussian pulse that leads to the Voigt CARS spectrum and, for this reason, provides the benchmark to which the new CARS lineshapes can be compared. In addition, we add a second symmetric pulse known in laser physics. It is the square of the hyperbolic secant (Sech2 in Fig. 1) [14, 15] and appears to characterize gain switching and mode locking in semiconductor systems [28, 29]. This pulse gives rise to a spectrum which is intriguing because of a strong affinity with the Gaussian spectrum. Indeed, the two coincide except for the slightly wider wings of the Sech2 spectrum. On the other hand, we are mostly concerned with time asymmetries and the square of the asymmetric hyperbolic secant (AsySech2 in Fig. 1) has attracted some attention to simulate realistic single-mode laser pulses [27, 30]. Finally, to introduce a further asymmetric pulse, we also use a combination of a parabola with a slower exponential decay (AsyExp in Fig. 1). This particular choice guarantees analytical results for the CARS signal and could be useful for practical applications where fast computing times are necessary.

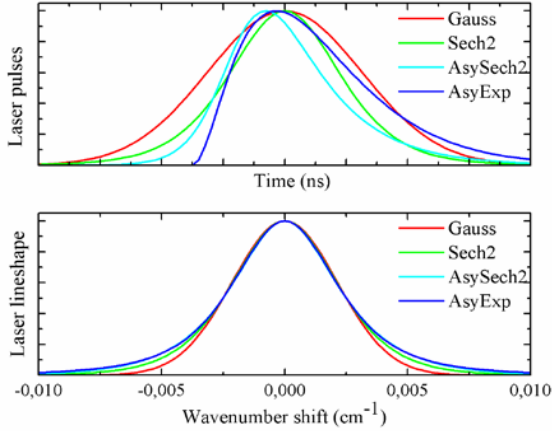

Fig. 1 (Upper panel) Shape of the laser pulses used in the analysis. Their time profile is constrained to the same FWHM of the corresponding lineshapes (lower panel).

To sum up, the four chosen laser pulses and their corresponding spectral representations are reported in Tab. 1 (apart from constant amplitudes that we choose in a manner that all the pulses have the same maxima in Fig. 1).

Table 1. Time and frequency representations of the laser pulses

| Time | Frequency |
| --- | --- |
| $e^{-\sigma_{Gauss} t}$ | $e^{-(\omega/\sigma_{Gauss})^2}$ |
| $\mathrm{sech}^2(\pi \sigma_{Sech2} t/2)$ | $\mathrm{sech}^2(\omega/\sigma_{Sech2})$ |
| $[2/(e^{\sigma_{AsySech2} t} + e^{-\alpha \sigma_{AsySech2} t})]^2$ | $abs^2[\sec g(\omega)]$ |
| $\vartheta(t) t^2 e^{-2\sigma_{AsyExp} t}$ | $abs^2[1/(i+\omega/\sigma_{AsyExp})^2]$ |

In the table, $g(\omega) = \pi[(1+3\alpha) - 2i\omega/\sigma_{AsySech2}]/[2(1+\alpha)]$ where the parameter $\alpha > 1$ is responsible for the asymmetry of the AsySech2 pulse (the condition $\alpha = 1$ corresponds to the Sech2 pulse). In Fig. 1, we set $\alpha = 3$ and this refers to a situation of a moderate asymmetry. Later, we will also consider more extreme conditions of stronger asymmetry by setting greater values of $\alpha$. Furthermore, another example of time asymmetry is reproduced by the introduction of the AsyExp pulse where an exponential decay follows a parabolic build-up. This particular time dependence is complemented by the theta function $\vartheta(t)$ that takes into account the requirement of causality (i.e., no Raman response is possible before the arrival time of the pulse). Finally, it is important to note that the chosen AsySech2 and AsyExp pulses are characterized by nearly identical spectral lineshapes.

Based on the description summarized in Fig. 1 and Tab. 1, we now compare the gas spectra for a common set-up where the CARS lineshape is dictated by the convolution between the narrower laser line and the Raman response of Eq. (1) [3, 6, 9-11, 13, 18, 20]. To this end, we have to set reasonable values of the Raman linewidth $\Gamma$ appearing in the nonlinear susceptibility and, since we deal with gases, it is appropriate to choose the interval within the extremes at 0.1 and 0.01 cm$^{-1}$. Physically, this range corresponds to moderate and low gas pressures with variable temperatures [3, 20, 31-34].

The results for the laser pulses of Fig. 1 are shown in Fig. 2. In the upper panel, the case of $\Gamma = 0.1$ cm$^{-1}$ is considered and, as expected, the CARS response to the different pulses is unaltered. In other terms, the CARS lineshape is the Lorentzian function of Eq. (1) and, as such, does not depend on the temporal or spectral details of the laser pulses. In this limit, it is thus justified the approximation of the spectral laser line to a delta function (CW limit of monochromatic laser).

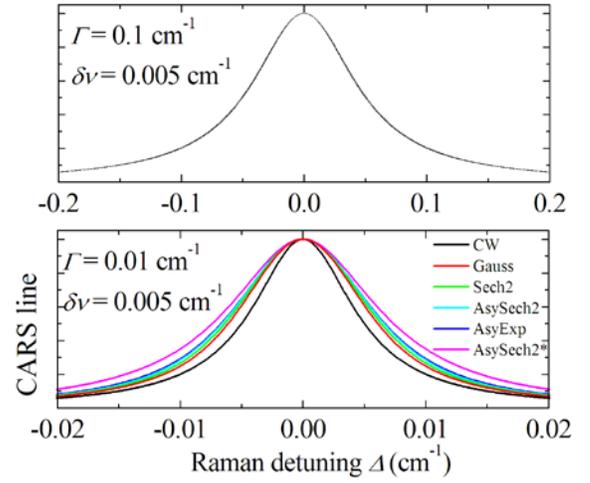

Fig. 2 CARS lineshapes related to the laser pulses of Fig. 1 having the same linewidth of 0.005 cm$^{-1}$. The upper and the lower panel refer respectively to the chosen extremes of 0.1 and 0.01 cm$^{-1}$ for the Raman linewidth.

The situation changes at the other extreme of $\Gamma = 0.01$ cm$^{-1}$. In this second condition, the details of the laser pulses are relevant. First of all, the Lorentzian limit of a CW monochromatic laser is now distinguishable from the Voigt profile corresponding to the Gaussian pulse (Gauss). If this is not surprising, it is however striking that the other laser pulses determine appreciably broader spectral shapes. The conclusion is already valid for the symmetric Sech2 pulse and is strengthened by the result obtained for the asymmetric pulses whose CARS lines are sufficiently different from the Voigt shape. In particular, the deviation from the Voigt dependence become clearer for a strongly asymmetric pulse (AsySech2* where $\alpha = 10$ in Tab. 1). The immediate consequence of this comparison is that the CARS approach exclusively based on the Lorentz or Voigt shape [6, 9-12, 18-20] introduces inaccuracies in relation to the role of the Raman linewidth $\Gamma$. This is exemplified in Fig. 3. Here, the relative variation $\delta\Gamma/\Gamma = 100(\Gamma_{Voigt}/\Gamma - 1)$ is plotted as though we were to use the Voigt CARS line to interpret data obtained with the other laser pulses.

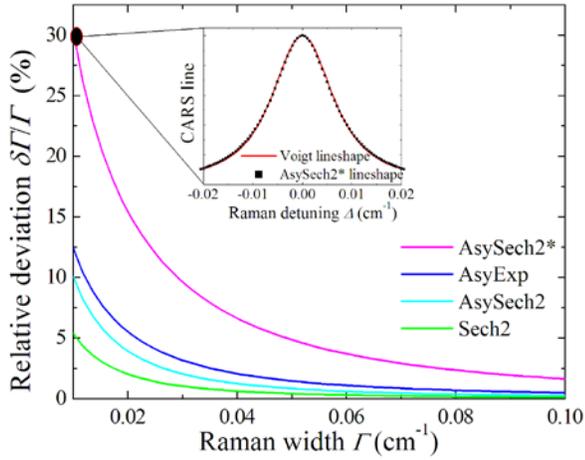

Fig. 3  Relative variation of the Raman linewidth as obtained from the Voigt lineshape used to reproduce the CARS lines obtained for the Sech2, AsySech2, AsyExp and AsySech2* laser pulses of Fig. 1. In the inset, the Voigt and the AsySech2* lines are shown for the largest deviation of 30 %.

One of the main results of Fig. 3 is that significant deviations at low $\Gamma$ values are already observed for the symmetric Sech2 pulse whose spectral difference from the Gauss pulse is limited to slightly longer wings (see Fig. 1). More importantly, in Fig. 3 the asymmetric pulses bring out larger deviations depending on the degree of time asymmetry that is reflected in the spectral convolution with the CARS susceptibility. It is then fitting at this juncture to conclude that the interpretation of CARS lineshapes has a critical stage in the deeper evaluation of the role played by the laser pulses.

To supplement this proof with a further evidence, we show next the analogous result of Fig. 2 by considering a larger spectral FWHM of 0.1 cm$^{-1}$ for the frequency representation of Tab. 1. In this case, we clearly disregard the relation with transform-limited laser pulses because of the very large laser bandwidth that points at multi-mode laser systems. Nonetheless, large laser bandwidths are typical in many studies and in many calculations of CARS lines [6-12, 20, 35]. For this reason, it is equally worthwhile to verify the modification of the CARS line in dependence of the spectral details of the laser pulses. The results are given in Fig. 4 assuming the same interval of variability for the Raman linewidth.

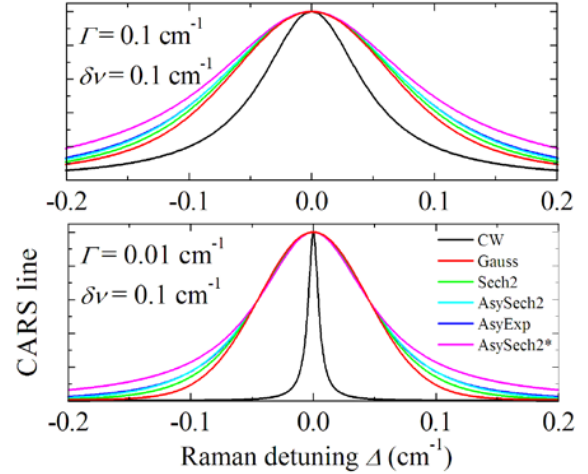

Fig. 4  CARS lineshapes related to the laser pulses of Tab. 1 constrained to the same spectral FWHM of 0.1 cm$^{-1}$. The upper and the lower panel refer respectively to the chosen extremes of 0.1 and 0.01 cm$^{-1}$ for the Raman linewidth.

In remarkable contrast with the result of Fig. 2, significant changes are already visible at the highest $\Gamma$ value (upper panel of Fig. 4). At the other extreme (lower panel) such changes remain so that the sensitivity of the CARS lineshape to the laser pulses is further confirmed.

The practical consequences of the analysis made in this Letter have to be emphasized in view of a better CARS accuracy, especially within the field of CARS diagnostics where the evaluation of the Raman linewidths plays a prominent part [3, 6, 8, 20]. As a matter of fact, gases at high temperature and moderate or low pressure are characterized by small collisional Raman widths for which the considerations raised by the results of this work are valid. For example, as noted by several authors [3, 6, 7, 20], a small error in the evaluation of the Raman linewidths produces a large error in the spectral analysis of CARS experiments. It is then not unexpected that authors, using the Gauss pulse leading to Voigt-type spectra, could find fatal thermometric mismatches at high temperatures [20]. Instead, according to the current work, the use of more realistic models for the laser pulses could correct such a limitation.

In the end, based on the foregoing considerations, we have demonstrated what follows.

(i) The CW limit of perfectly monochromatic laser pulses, sometimes adopted under single-mode operation, fails at high temperature and moderate (or low) pressure.

(ii) Transform-limited pulses with a time symmetry similar to the Gauss pulse show some differences in the comparison with the Voigt lineshape.

(iii) Time-asymmetric transform-limited laser pulses generate CARS lines that deviate even more from the Voigt profile.

(iv) There exist time-asymmetric pulses that can be modeled by expressions useful for analytical

representation of the CARS line in place of the Voigt function (here, the AsyExp pulse leads to the line $\text{Im}\{\pi\sigma_{AsyExp}[\Delta - i(\Gamma/2 + 2\sigma_{AsyExp})]/2[\Delta - i(\Gamma/2 + \sigma_{AsyExp})]^2\}$

(v) CARS lineshapes at any reasonable value of the Raman linewidth are particularly sensitive to laser pulses that are not transform limited.

To conclude, we have introduced new lineshapes of nanosecond CARS in relation to realistic properties of the laser pulses. This leads to a revision of the role played by the Raman linewidths when these are evaluated in the known CARS theory or in numerical codes based on the Gaussian laser bandwidth. The results of this study provide then a rigorous perspective on what suggested a long time ago by Farrow et al. as to comparisons between CARS theory and measurements: "better agreement would require more detailed modeling of the laser line shapes, such as the use of measured spectral profiles in the convolution calculations rather than the Gaussian line shape approximations" [36].